\documentstyle[12pt,aasms]{article}
\begin{document}

\title{Analysis of low z absorbers in the QSO spectra}

\author{R. Srianand\\
        Inter-University Centre for Astronomy and Astrophysics \\
        Post Bag 4, Ganeshkhind, Pune 411 007, India,\\
	email anand@iucaa.ernet.in}

\begin{abstract}

We present the results of reanalysis of low redshift Mg II absorption
line sample compiled by Steidel and Sargent. We constructed grids of
photoionization models for various cloud parameters. The conditions on
cloud parameters to produce N(Fe II)$\ge$N(Mg II) are obtained using
single cloud curve of growth. Properties of Mg II absorbers with W(Fe
II)/W(Mg II)(definded as R) $\ge$ 0.5  and with R$<$0.5 are analysed
separately. Contrary to the whole Mg II sample, the clouds with R$<$0.5
show a steep increase in number density with redshift.  These systems
also show clear increase in W(Mg II) and doublet ratio of Mg II with
redshift. However there is no correlation between W(Mg II) and doublet
ratio. In the case of R$\ge$0.5 clouds W(Mg II) and doublet ratio are
not correlated with redshift. However there is a clear anticorrelation
between doublet ratio of Mg II and W(Mg II). We find a clear decrease
in the ratio of W(Fe II 2383) and W(Mg II 2796) with redshift. The
number density of Fe II lines selected absorbers are not evolving with
redshift, consistent with Mg II results. We are also not finding any
dependence of W(Fe II 2382) and the ratio of W(Fe II 2382) and W(Fe II
2600) with redshift. This implys no evolution for the average Fe II
column density with redshift. Based on the available data of Lyman
limit sytems (LLS) in the literature, we are not finding any dependence
of optical depth ($\tau_{\rm LLS}$) on redshift in the range z =
0.3-2.0. We collected the LLS information for 53 QSO sight lines, from
the literature, for which details of Mg II absorption are available.
There are 4 Mg II absorption systems which are not LLS at redshifts
which are lower than the mean redshift of the sample (${\rm
z\simeq1.1}$). In the higher redshifts, where one expect to see
2.5$\pm$1.4 such absorbers, we do not find any nonLLS Mg II absorbers.

Individual systems with ${\rm \tau_{LLS}<3.0}$ are analysed with an aim
to  constrain the ionization parameter and metallicity.  Our results
imply some of the absorbers at z$\simeq$0.6 have reached metallicity
roughly around solar value, indicating the chemical enrichment in some
of the absorbers are similar to our Galaxy, as z$\sim 0.6$ is roughly
the formation epoch of sun in our Galaxy. The required ionization
parameters for these systems are less than 0.001 in most cases.
Comparison of our results with results obtained for intermediate and
high redshift absorbers confirm that mean ionization state of metal
rich  absorbing clouds falls with redshift.

\vskip0.4in \noindent {\it Subject headings:} QSO: absorption
lines-QSO:general \end{abstract}

\section{INTRODUCTION}

Absorption lines seen in the spectra of QSOs provide an indirect mean
by which the formation and evolution of galaxies can be studied.
Latest observations of bright galaxies at the redshifts of the
previously known Mg II absorbers (Bergeron (1988); Steidel (1993))
strengthen the idea that the absorbers are the gas clouds present in
the extended halo regions of the luminous galaxies. Results obtained by
studying the redshift evolution of these absorbers can be utilized to
understand the evolution of the gas content of these galaxies.

Properties of the metal line absorption systems can be analysed with
photoionization models to get the estimates of metal abundance and
intensity of metagalactic ionizing flux. (eg. Bergeron and Stasinska
1986; Srianand and Khare 1994, etc..). These studies use high redshift
observations to get constraints on various model parameters, as most of
the ultra violet transitions produced at higher redshifts alone can be
observed using ground based telescopes.

In the low z range (i.e z = 0.2-2.0) Mg II lines together with Fe II
lines fall in the observable window of the ground based telescopes. In
order to get other UV transitions of the abundant elements, in this
redshift range, one has to go for space telescope observations. Another
way of understanding the nature of the low z absorbers is to image and
study the properties of actual galaxies which are responsible for
absorption.  Steidel(1993) and his collaborators surveyed all the
galaxies that are responsible for Mg II absorption (for z$<1.5$). Based
on optical and IR imaging and spectroscopy they concluded that the
galaxies producing Mg II absorption lines are similar to the normal
galaxies at the present epoch. They did not find any evolution in B-K
colour, space density and luminosity of the galaxies selected based on
Mg II absorption.  Their study showed on an average Mg II absorbing
galaxies appear to be consistent with a normal 0.7 ${\rm {L_B}^*}$  Sb
galaxy having a roughly constant star formation rate since z$\simeq$1.

It is known that the look back time corresponding to the epoch of
formation of our sun is z $\simeq$0.692(for ${\rm H_o =
75\;kms^{-1}\;Mpc^{-1}}$ and ${\rm q_o =0.5}$ with the age of the sun
taken to be 4.6 Gyr). Thus if the absorbing galaxies are similar to
normal galaxies, one would expect to see some of the absorbers to have
metal abundances almost that of sun. If the QSOs and other AGNs are the
main contributers of the intergalactic ionizing UV background then the
ionizing flux at the Lyman limit is expected to decrease with
decreasing redshift.  (Miralda-Escude and Ostriker 1990; Madau 1992)
and one would expect to see more domination of low ionization lines at
low redshifts.

Statistical stydies of Mg II absorption lines by Steidel and Sargent
(1992), with their homogeneous high signal to noise data, show there is
a differential evolution in the  number density of Mg II absorbers with
redshift. Strong lines show more redshift evolution compared to weak
lines. This together with the nonevolving Mg II doublet ratio are
interpreted as the evolution in number of subcomponents producing the
Mg II absorption lines.

Recent HST observations indicate the decrease in number density of C IV
systems with decreasing redshift (for z$<$ 1.0)(Bahcall etal 1993).
This together with the finding of low neutral hydrogen optical depth
low ionization systems (Bergeron et al 1994) show that the mean
ionization of the metal rich optically thin absorbing clouds falls with
the redshift.

In this work we study the properties of low z absorbers using the data
of Steidel and Sargent (1992) considering the ratio of equivalent
widths of Fe II(2382) and Mg II (2796) lines rather than considering Mg
II lines alone. We have performed several grids of photoionization
models and using standard single cloud curve of growth we obtained
conditions for clouds to have various equivalent width ratios of Fe
II(2382) and Mg II(2796)  lines, which are described in
section 2. In section 3 redshift evolution in number density of Mg II,
Fe II lines and LLS are analysed. In section 4 we present the results
of correlation tests performed on equivalent width distributions. In
section 5 we analyse the properties of LLS which are also having Mg II
observations. Discussion and results are presented in section 6.

\section{CURVE OF GROWTH AND PHOTOIONIZATION MODELS}

We performed standard single cloud curve of growth to calculate the
column densities of Fe II, for various values of N(Mg II), required to
produce different values of R.  Figure 1 shows the expected
relationship between Fe II and Mg II column densities for different
values of equivalent width ratios and velocity dispersion parameters.
We used the atomic parameters given in Morton (1992) for these
calculations. It is clear from the figure that the N(Fe II) will be
greater than or equal to N(Mg II) when the R is greater than 0.5.

One can get the conditions on cloud parameters, to produce N(Fe II)
$\ge$ N(Mg II), by constructing photoionization models. We constructed
several grids of photoionization models using the code "CLOUDY" kindly
given to us by Prof. Ferland. In these calculations we assumed the
cloud to be a plane parellel slab, photoionized by the powerlaw
background radiation.  The ratios of various elements are assumed to be
solar ratios. The calculations are performed for various values of
ionization parameter, $\Gamma$, metallicity, Z, and neutral hydrogen
column density. Figure 2 shows the N(Fe II)/N(Mg II) as a function of
neutral hydrogen column density for various cloud parameters.  It is
known from the standard curve of growth that inorder to produce
observable equivalent width (i.e  $>0.2-0.3\AA$) the Fe II column
density should be greater than 10$^{13}\;{\rm cm^{-2}}$.

In table 1 we give the minimum neutral hydrogen column densities needed
to produce N(Fe II) $>10^{13}\;{\rm cm^{-2}}$ for various cloud
parameters.  Also given in the table are the minimum values of N(H I)
to produce N(Fe II)$\ge$N(Mg II), and the expected minimum Mg I column
density for this value of N(H I). It is clear from the table that the
cloud has to be a damped Lyman alpha cloud, when ($\Gamma$) is
greater than 0.001, inorder to produce R $>$ 0.5, when Z$<{\rm
0.5Z_\odot}$. Also Fe II lines will be observable only in the case of
high optical depth Lyman limit clouds.  If the value of ($\Gamma$) is
less than 0.001, higher optical depth LLS as well as damped Lyman alpha
clouds will produce R $>$ 0.5.  Thus if the cloud is photoionized by
the background radiation then it can produce R $>$ 0.5 only when the
optical depth of hydrogen is very high or the metallicity is very high
when N(H I) $<\;10^{19}\;{\rm cm^{-2}}$.  Also in these high
metallicity cases Fe II lines can be produced even by the clouds with
hydrogen Lyman limit optical depth less than 1.

We know that oscillator strength of Mg I is large and a cloud can
produce a line with W(Mg I 2852)$>$ 0.3 when  N(Mg I) $> 10^{12.5}{\rm
cm^{-2}}$. Thus from the table  it is clear that for  ${\rm
(\Gamma)}>0.001$ whenever R $>0.5$ one should see Mg I(2852) line
also.  Thus absence of Mg I line when R $>0.5$  will indicate low
neutral hydrogen optical depth and hence low ionization parameter.

Thus it is clear, from this analysis, that studying the evolution of
W(Fe II 2382)/W(Mg II 2796) with redshift will enable one to study the
evolution of ionization parameter and neutral hydrogen optical depth.
Motivated by this idea in the following sections we perform various
statistical tests using the Mg II sample compiled by Steidel and
Sargent (1992) and low z LLS samples (Lanzetta, Wolfe, Turnshek, 1995;
Bahcall et al.  1993).

\section{NUMBER DENSITY DISTRIBUTION OF ABSORBERS}

It is customary to parameterise the redshift distribution of absorption
line systems as

\begin{equation}
{\rm N(z) = N_o(1+z)^{\gamma}},
\end{equation}
where ${\rm N_o}$ and $\gamma$ are constants that are to be determined
by fitting the observed distribution. Steidel and Sargent (1992) used
maximum likelihood analysis to calculate the values of ${\rm N(z)}$ and
$\gamma$ for various subsamples. Their results are given in table 2.
For the same subsamples, considering same redshift path for each QSO,
we have calculated the value of $\gamma$ and ${\rm N(z)}$ considering
only lines with R$< 0.5$. This based on our curve of growth results
implies that  N(Fe II) to be less than N(Mg II).  Whenever the Fe II
equivalent widths are not available we have taken the 5$\sigma$
uncertainties as the upper limits.  The values obtained using the
maximum likelihood analysis are given in Table 2. It is clear from the
table that, contrary to the whole Mg II sample considered by Steidel
and Sargent (1992), Number density of absorption systems with total Mg
II column density greater than total Fe II column density show a sharp
increase with redshift. This implies at higher redshifts, most often,
the Mg II lines tend to be much stronger compared to Fe II lines.

We considered all the Fe II lines irrespective of Mg II equivalent
widths and performed maximum likelihood method to get the evolutionary
index.  In these calculations we used only regions where one can detect
Fe II lines with equivalent width greater than the  cutoff equivalent
width used for different samples. The results obtained are also given
in table 2. Like Mg II sample the results are consistent with
nonevolving absorbers for 0.0$\le{\rm q_o}\le$0.5. However the small
number of Fe II lines involved in these calculations prevent us from
making any conclusions regarding differential redshift evolution.

IUE observations of low redshift Lyman limit systems are available in
Lanzetta, Wolfe, Turnshek(1995). We combined their data set with the
HST observations obtained from the literature (Bahcall et al. 1993) and
performed maximum likelihood analysis to study the redshift evolution
of LLS in the low redshifts (i.e $<2.0$). The observed value of
$\gamma$ for the whole sample with ${\rm \tau_{LLS}>1.0}$ is
1.30$\pm$0.65. When we consider only strong LLS with ${\rm
\tau_{LLS}>3.0}$ the obtained value of $\gamma$ is 1.56$\pm$0.84 and
$\gamma$ is 0.90$\pm$1.30 for LLS with ${\rm \tau_{LLS}<3.0}$. There
seems to be no change in the value of $\gamma$ with the strength of the
LLS. The changes, if at all, are with in 1$\sigma$ uncertainity. Thus
the available data is consistent with no change in the Lyman limit
optical depth with redshift, in the redshift range 0.2 to 2.0. We
performed generalised Kendall rank correlation test (described in
Isobe, Feigelson $\&$ Nelson, 1986) to see any correlation between
${\rm \tau_{LLS}}$ and redshift. The null hypothesis that the two
quantities are uncorrelated can be rejected only at 55$\%$ confidance
level. Thus optical depth of the neutral hydrogen in LLS is not seem to
be correlated with redshift.

\section{REST EQUIVALENT WIDTH DISTRIBUTION}

The equivalent width distribution of any absorption line
can be parameterised as
\begin{equation}
 {\rm f(W)dW = ({N_*\over W_*})exp(-W/W_*)dW},
\end{equation}
where the parameters ${\rm N_*}$ and ${\rm W_*}$ are determined using
maximum likelihood analysis. The values of ${\rm W_*}$ for Mg II lines
with R$>0.5$  are consistent with the ${\rm W_*}$ values of Mg II
absorption lines obtained for the whole sample(table 2). This implies
that presence of Fe II lines with equivalent width greater than
$0.5\times {\rm W(Mg II)}$ is not correlated with the equivalent width
of the Mg II lines. The values of ${\rm W_* } $ obtained for Fe II
lines are less than that for Mg II systems. This clearly shows that the
average Fe II equivalent width is less than that of the Mg II
equivalent width.

It is known for the case of Mg II systems, that the average equivalent
width increases with increasing redshift (Steidel and Sargent, 1992).
We performed Spearman correlation test to search for any possible
dependence of W(Fe II 2382) on redshift. The result is  given in table
3.  Contrary to the Mg II systems there is no correlation between W(Fe
II) and redshift, and the null hyphothesis that the two quantities are
uncorrelated can be rejected only at 37$\%$ confidence level. We also
do not find any correlation between W(Fe II 2382)/W(Fe II 2600) and
redshift. This implies that total column densities of Fe II lines are
not changing with redshift.  This with nonevolving Mg II doublet ratio
(dr) with z suggest that the relative column density of Mg II and Fe II
lines are not changing with redshift. This result also implies the
decrease in the equivalent width ratios of W(Fe II 2382) and W(Mg II
2796) with redshift.

In fig 3. we have plotted R as a function of redshift.  The open
squares in the figure represent the observed values and the closed one
represent the upper limits. The lines with R greater than 0.6$\AA$ are
rare in the higher redshift (say z$>$1.1). To illustrate this point we
have plotted the histogram of R in two different redshift bins (Fig
4).  It is clear from the figure that the distribution of R is changing
with redshift. The Spearman correlation test shows a 1.9$\sigma$
anticorrelation between R and redshift. If we include the upper limits
also  the null hyphothesis that the two quantities are uncorrelated can
be rejected at 94$\%$ confidence level using the generalised Kendall
rank correlation tests.  Thus various analysis performed with the
Steidel and Sargent (1992) data indicates an increase in the number of
lines, with  R $<$ 0.5, with increasing redshift.

We looked for the correlation between doublet ratio of Mg II lines and
R. This is similar to looking for the correlation between the column
density of Mg II lines and the column density ratios of Fe II and Mg II
lines. We do not find any correlation between the two quantities and
the null hyphothesis that the two quantities are uncorrelated can be
rejected only at 49$\%$ confidence level. We also do not find any
correlation between W(Fe II 2382)/W(Fe II 2600) and Mg II doublet
ratio. The null hyphothesis that the two quantities are uncorrelated
can be rejected only at 72$\%$ confidence level. This is consistent
with the nonevolving Mg II doublet ratio and the ratio of W(Fe II 2382)
and W(Fe II 2600) with redshift.

The change in the Mg II equivalent width with z is mainly interpreted
as due to the change in the number of components comprising the
line(Sargent and steidel 1992). Our results indicate that there is no
change in the Fe II column density with redshift. Thus the change in R
can be understood as the change in the relative number of clouds
producing Fe II and Mg II lines. While number of clouds (i.e. W(Fe II))
and the average column density of Fe II are not changing with redshift,
the number of clouds producing Mg II changes with redshift.  As the
presence of Fe II with Mg II lines imply the presence of high ${\rm
\tau_{LLS}}$ then this results imply that there are more highly ionized
component compared to low ionization components at higher redshift.
Thus in all cases when W(Fe II 2382) is much less than W(Mg II) one
should see W(C IV)$>$W(Mg II). Table 6 of steidel and Sargent (1992),
with the available Fe II observations, revel that in most of the cases
when W(C IV ) is greater than W(Mg II) the Fe II lines are absent.

Another surprising result of the analysis of steidel and Sargent (1992)
is the significant anti-correlation between Mg II doublet ratio and
W(Mg II 2796), at approximately 98$\%$ confidence level. This
correlation is paradoxical considering the fact that W(Mg II 2796) is
increasing with increasing redshift and doublet ratio of Mg II is
independent of redshift. They argued that this apparent correlation is
mainly due to the strong lines which tend to have doublet ratios near
one, while the weak lines have a wide range of doublet ratios.  In
order to understand this correlations we performed rank correlation
test between different parameters considering the Mg II lines with
R less than 0.5  and greater than 0.5 separately.  The
results of Spearman correlation tests are given in table 3.

For  systems with R${> 0.5}$ there is no strong tendency of Mg II
equivalent width as well as Mg II doublet ratio to show any change with
redshift. This result is consistent with the results obtained for Fe II
lines. However there is a clear anti-correlation between Mg II
equivalent width and doublet ratios. Thus the very high optical depth
systems do not show any change in Mg II or Fe II column density with
redshift.  Most probably the Mg II equivalent width is a measure of
column density rather than the number of clouds in these systems.

For   systems with  R${< 0.5}$ there is a
clear tendency of equivalent width of Mg II lines to increase with
redshift. There is also a weak correlation between Mg II doublet ratio
with redshift. The null hypothesis that the two quantities are
uncorrelated can be rejected at roughly 93$\%$ confidence level. This
indicates the decrease in total column density of Mg II with increasing
redshift. This results is similar to the one seen for C IV absorbers
for z$>$2.0. We do not find any significant correlation or
anti-correlation between Mg II doublet ratios and equivalent widths.

Thus in the total Mg II sample the correlation between W(Mg II) and
redshift is mainly due to clouds with R is less than 0.5, which are
also showing increase in Mg II doublet ratio with redshift. The
correlation seen between W(Mg II) and doublet ratios of Mg II is due to
clouds with R$>0.5$, which are not showing any correlation between W(Mg
II) and z, and redshift and doublet ratio of Mg II lines. In this way
we can understand the paradoxical correlation seen by steidel and
Sargent (1992).

\section{LOW REDSHIFT LYMAN LIMIT SYSTEMS}

There are 53 sight lines for which Mg II observations as well as low
redshift Lyman limit observations are avialable in the
literature(Steidel $\&$ Sargent (1992); Aldcroft, Bechtold $\&$ Elvis
(1994); Boisse et al.  (1993); Lanzetta, Wolfe $\&$ Turnshek (1995)
$\&$ Bahcall et al. (1994)).  Twenty eight LLS are with
$\tau_{\rm LLS}>1.0$ are detected in these spectra. Out of these 28
systems 7 systems do not show Mg II or Fe II absorption lines and rest
of them show metal line absorption at the same redshift. Four Mg
II systems which are not LLS are also in our sample. Out of 19 Mg II
systems which are LLS 5 systems have observed $\tau_{\rm LLS}<3.0$.
Thus Lyman limit optical depth is known for these systems and presence
of Fe II lines with Mg II lines in these low optical depth systems can
be used to put bounds on the physical conditions.

All the nonLLS systems in our sample are at redshifts which are less
than the average redshift of the sample (i.e z = 1.1). We calculated
the number density per unit redshift of the nonLLS absorbers, for
z$<$1.1, to be 0.28.  If we assume the number density of such absorbers
are not changing with redshift one should expect to see 2.5$\pm$1.4
absorbers for ${\rm z>1.1}$ in our sample.  Our failure to detect any
such absorbers at high redshift suggests the decrease in the ionization
parameter and/or increase in the metal abundance in the Mg II absorbers
with decreasing redshift. The number per unit redshift of LLS which are
not showing Mg II or Fe II lines in the two redshift bins are
0.33$\pm$0.16 and 0.32$\pm$0.19 respectively. The values are consistent
with one another suggesting no evolution for such systems with
redshift. If these LLS are highly ionized metal line systems, (i.e with
observable C IV), then our results suggest that very high ionization
metal poor clouds are present even at low redshifts.

As seen in the previous section the presence of Fe II with
the Mg II lines in the low neutral hydrogen optical depth clouds will
reflect very high metallicity and low ionization parameter. In what
follows we try to get the column densities of metal lines assuming
single cloud curve of growth. Then these column densities together with
the results of photoionization models are used to get an estimate of
the metal abundance and the ionization structure of the clouds.

As we know almost all the metal lines are produced by the ensemble of
clouds rather than by a single cloud as assumed here. However Jenkins
(1986) showed that if the components forming the blends are not heavely
saturated, the column density obtained using doublet ratios and single
cloud curve of growth will be equal to the actual total column density
and the obtained effective velocity dispersion will reflect the number
of clouds rather that the kinematics of the clouds. Also in the
photoionization models, we are using the shape of the ionizing
background to be a single powerlaw.  In the realistic case the
continuum will be altered by the intervening material in the
intergalactic medium. However in the redshift range we are interested
in, the absoption due to intergalactic material will be less
pronounced and our assumption of powerlaw will be a reasonable one.
Thus our analysis will not give the exact values of the metallicity and
background radiation however one can get a bound on these values.

In what follows we analysis the 9 low optical depth systems one by one
to get a rough estimate of the background radition and the
metallicity.

\subsection{Systems with $\tau_{\rm LLS}<1.0$}

\noindent(a) 1115+080 (${\rm z_{abs}\simeq\;1.0430}$)

IUE spectrum of this QSO is clean (Lanzetta, Turnshek $\&$ sandoval,
1992) and does not show any break in the expected position. Steidel and
Sargent (1992) indentified this weak Mg II system. The rest equivalent
widths of the Mg II doublets are 0.31 and 0.18 $\AA$ respectively. The
high doublet ratio indicates that Mg II lines are in the linear portion
of the curve of growth. Single cloud curve of growth gives N(Mg II) =
9$\times 10^{12}\;{\rm cm^{-2}}$ with the effective velocity
dispersion, ${\rm b_{eff}\;=\;40\; km\;s^{-1}}$.  The absence of break
in the continuum at the expected position of the Lyman limit gives an
upper limit on the neutral hydrogen column density to be 8$\times
10^{16}\;{\rm cm^{-2}}$ ($\tau_{\rm LLS}<0.5$).  Young, Sargent and
Boksenberg (1982) did not detect Fe II lines for this system.  The
observed 5$\sigma$  upper limits on the Fe II(2382) and Fe II(2600)
lines (0.10 and 0.15 $\AA$ respectively) give the value of N(Fe II)$\le
7\times 10^{12}\;{\rm cm^{-2}}$ for ${\rm b_{eff}\;=\;40\;
km\;s^{-1}}$. Upper limit on Mg I line give N(Mg I) $< 10^{12}\;{\rm
cm^{-2}}$.  Absence of Fe II lines and weak Mg II unsaturated lines
prevent us from getting any stringent constraints on the radiation
field and metalicity. For the metallicity to be around 0.1${\rm
Z_\odot}$ the ionization parameter for this system should be less than
$10^{-3}$. Along the same line of sight, Young, Sargent and Boksenberg
(1982) observed five other absorption systems ( at redshifts 1.6998,
1.7283, 1.7304, 1.7322, and 1.7535). No Mg II absoption lines are seen
for these systems (Steidel $\&$ Sargent 1992). This with nondectection
of LLS show that these systems are highly ionized with ionization
paramter greater than $10^{-2}$.

\noindent(b) 1206+457 (${\rm z_{abs}\simeq\;0.9277}$)

IUE spectrum of this QSO shows no sign of break in the expected
position (Lanzetta, Turnshek $\&$ sandoval, 1992).  However  a very
strong Ly $\alpha$ with rest equivalent width 7.5 $\AA$ is observed for
this systems (Lanzetta, Wolfe, Turnshek 1995). This clearly indicates a
complex blend with very high effective velocity dispersion (Since N(H I
) is small). Mg II lines were identified by Steidel and Sargent
(1922).  Mg II lines are also strong (with equivalent widths 1.0 and
0.79 $\AA$) and saturated. The single cloud curve of growth gives the
total Mg II column density to be greater than  7$\times 10^{13}\;{\rm
cm^{-2}}$ for ${\rm b_{eff}\;=\;40\; km\;s^{-1}}$. Upper limits on the
Mg I equivalent width gives N(Mg I) $<10^{12}\;{\rm cm^{-2}}$. Fe
II(2382) line is observed with rest equivalent width 0.27 $\AA$. This
with the $5\sigma$ upper limits on the equivalent width of Fe II(2600)
line give N(Fe II) $\simeq$ 2$\times10^{13}\;{\rm cm^{-2}}$. These
values imply the ionization parameter to be less than 3$\times10^{-4}$
and the metallicity to be greater than 0.5$\times{\rm Z_\odot}$. Very
strong Ly $\alpha$ line with $\tau_{\rm LLS}<1$ indicates Ly $\alpha$
effective velocity dispersion to be more than that for Mg II lines.
This imply that there are Ly $\alpha$ clouds which do not produce
appreciable amount of Mg II lines. If the absorbing region consists of
two components, one is highly ionized and the rest is a low ionized
region, then most of the Ly $\alpha$ absorption seen in this system may
be due to the high ionization phase.  In such case HST observations
should show a broad C IV line for this system. If the observed C IV
lines are also narrow then the high Ly $\alpha$ equivalent width will
reflect the clustering of Ly $\alpha$ forest lines with this metal line
system.

\noindent(c) 1338+416 (${\rm z_{abs}\simeq\;0.6213}$)

This system is analysed in detail by Bergeron et al (1995). This
system has R$\simeq$0.5, which implies N(Fe II) $\simeq$ N(Mg II).
This with $\tau_{\rm LLS}<1.0$ indicate high metallicity and low
ionization for this system.

\noindent(d) 1040+122 (${\rm z_{abs}\simeq\;0.6591}$)

Aldcroft, Bechtold, $\&$ Elvis (1994) reported this systems for the
first time. However they suggested the identification to be marginally
significant with the disagreement in the redshift matching of the
doublet around 0.0006. The fit to the Mg II doublets showed a clear
wavelength ratio problem, however the fit had ${\chi_\nu}^2$ =1. If
this identification is real then this is also a system with  very low
Lyman limit optical depth. Bahcall etal. (1993) do not find any break
in the expected position of this redshift. The equivalent width of Mg
II doublets are 0.58 and 0.42 $\AA$ respectively. Mg I line is not
observed and the 5$\sigma$ upper limit is 0.19$\AA$.  Fe II lines are
absent and the observed 5$\sigma$ limit on the equivalent width is
0.40$\AA$. The single cloud curve of growth gives the N(Mg II)$\ge
3\times10^{13}{\;\rm cm^{-2}}$. This high column density of Mg II with
low values of ${\rm \tau_{\rm LLS}}$ imply ionization parameter
$\Gamma<0.004$ and metallicity $>0.1{\rm Z_\odot}$

\subsection{$1<\tau_{\rm LLS} <$3 systems}

\noindent(a) 1017+280 (${\rm z_{abs}\simeq\;1.9230}$)

This system has $\tau_{\rm LLS}\simeq $1.0 (Lanzetta, Turnshek $\&$
Wolfe(1995), which implies N(H I) = 1.6$\times 10^{17}\;{\rm
cm^{-2}}$.  Fe II lines are not observed for this system. C IV doublet
lines are observed for this system with equivalent widths 0.37 and 0.25
$\AA$ implying N(C IV) $\simeq2\times10^{14}\; {\rm cm^{-2}}$ and
b$_{\rm eff}\simeq30\; {\rm km\; s^{-1}}$. Based on the observed C II
lines one can get N(C II) $\simeq 6\times 10^{13}\;{\rm cm^{-2}}$.
Absence of Si II lines give an upper limit in the column density N(Si
II) $\le 10^{12}\;{\rm cm^{-2}}$. Si IV lines are observed for this
system and single component curve of growth gives N(Si IV) $\simeq
4\times 10^{13}\;{\rm cm^{-2}}$. Steidel and Sargent (1992) observed
very weak Mg II lines, with equivalent width 0.11 and 0.12$\AA$
respectively.  This gives Mg II column density to be 8$\times
10^{12}{\;\rm cm^{-2}}$.  Absence of Fe II lines with W(Si IV)$>$W(Si
II) and W(C IV)$>$W(C II)$>$W(Mg II) imply a high ionization. This
system is analysed in detail with a more realistic multicomponent
photoionization models by Srianand $\&$ Khare (1994). Their results
show (see their table 5) the required ionization parameter for this
systems is $\simeq10^{-2}$ and the average metallicity of the clouds
producing the absorption is roughtly around 0.014${\rm Z_\odot}$.

\noindent(b) 1317+277 (${\rm z_{abs}\simeq\;0.6601}$)

This system has $\tau_{\rm LLS}\simeq 2.7$ (Lanzetta, Wolfe $\&$
Turnshek 1995) implying that the N(H I) = 4$\times 10^{17}\;{\rm
cm^{-2}}$. The equivalent widths of the Mg II doublets are 0.49 and
0.33 $\AA$ respectively. The strong Mg II lines with doublet ratio
$\sim1.5$ give N(Mg II) $>2.2\times10^{13}\;{\rm cm^{-2}}$ for the
effective velocity dispersion parameter 25 kms$^{-1}$. Fe II lines are
observed with W(Fe II 2382) as strong as that for Mg II lines (0.42
$\AA$) indicating N(Fe II) $>$ N(Mg II). This together with low neutral
hydrogen optical depth imply the ionization parameter to be less than
3$\times 10^{-4}$. The metallicity in this absorber may very well be
higher than solar metallicity.

\noindent(c) 1354+19 (${\rm z_{abs}\simeq\;0.4591}$)

This system is discussed in detail in Bergeron et al.(1995) This system
has $\tau_{\rm LLS} = 1.2 \AA$. The equivalent widths of the Mg II
doublets are 0.89 and 0.82. The Mg II line is saturated and N(Mg
II)$>4\times 10^{13}{\rm cm ^{-2}}$. The observed equivalent width of
Fe II(2600) = 0.32$\AA$ and the 5$\sigma$ upper limit on W(Fe II 2382)
implies saturation for Fe II lines and the N(Fe II) values for this
will be $>1.5\times 10^{14}{\rm cm^{-2}}$.  The Mg I line is observed
with equivalent width 0.16 $\AA$. Assuming Mg I to be optically thin
one can get N(Mg I) = 1.5$\times 10^{12}{\rm cm^{-2}}$.  High values of
Mg II and Fe II column densities together with low N(H I) imply
metallicity above solar and very low ionization parameter, less than
($10^{-4.0}$). Bechtold $\&$ Ellingson (1992) found the galaxy which is
responsible for producing the Mg II absorption lines.  The impact
parameter with the QSO line of sight from the system is
27.2h$_{100}^{-1}$  kpc$^{-1}$.

\noindent(d) 1634+706 (${\rm z_{abs}\simeq\;0.9903}$)

This system has $\tau_{\rm LLS}\simeq 1.1$ which implies N(H I) =
1.76$\times10^{17}\;{\rm cm^{-2}}$. The observed Mg II equivalent
widths for this system is 0.58 and 0.42 $\AA$ respectively. Fe II and
Mg I lines are not detected. Single component curve of growth gives
N(Mg II) $\ge3\times 10^{13}{\rm\;cm^{-2}}$. The lack of Fe II lines
implies the ionization parameter in the range -3.0 to -4.0 for the
metallicity greater than 0.1${\rm Z_\odot}$.

\noindent(e) 1038+063 (${\rm z_{abs}\simeq\;0.458}$)

This system is analysed by Bergeron et al (1995) in detail. This system
has ${\tau_{\rm LLS}=1.2}$ with a saturated Mg II doublet similar to
${\rm z_{abs}\simeq\;1.0384}$ absorber in the spectra of 1634+706. This
system also implies metalicity more than 0.5 ${\rm Z_\odot}$ and
ionization parameter in the range $10^{-3.0}$ to $10^{-4}$ (Bergeron et
al.  1995). The galaxy causing these absorption lines is identified at
a projected radial seperation of 35${\rm {h_{100}}^{-1}}$ kpc from the
QSO sight line (Bergeron $\&$ Boisse 1991).

It is known that most of the iron in our interstellar medium is in the
form of dust. Thus if the absorbers discussed here have similar dust
content then the obtained metallicity estimates are very well be lower
limits.

\section {Results and discussion}

(1) We performed standard single cloud curve of growth to calculate
column densities of Fe II required to produce different values of R. It
is shown that when R $\ge$ 0.5, the  column density of Fe II will be
greater than equal to column density of Mg II.  Simple photoionization
models show, such absorbers are possible only in the damped Lyman alpha
clouds when the ionization parameter is more than 0.001.  If the
absorbers are with low Lyman limit optical depth (say $\tau_{\rm LLS}<
3.0$) then the presence of Fe II lines, with R greater than equal to
0.5, will imply very low ionization parameter and high metallicity.

(2) The redshift distribution of Mg II, Fe II and Lyman limit systems
are analysed using maximum likelihood analysis. Contrary to whole Mg II
sample considered by Steidel and Sargent (1992) the number density of
absorption systems with R$<$0.5 are showing clear increase in number
density with redshift. The redshift distribution of Fe II lines are
consistent with nonevolving population of absorbers. The number
densities of low z LLS are also consistent with the no evolution in
redshift. The evolutionary index $\gamma$ is independent of the optical
depth cutoff.

(3) Equivalent width distribution of subsampels of Mg II systems are
analysed using maximum likelihood analysis. The values of ${\rm W_*}$
obtained for lines with R greater that 0.5 are consistent with the
${\rm W_*}$ values obtained for whole Mg II sample. This implies the
presence of strong Fe II lines are not correlated with equivalent width
of Mg II lines.

(4) We do not find any trend of W(Fe II 2382) and the ratio of W(Fe II
2382) and W(Fe II 2600) with redshift.

(5) There is a clear increase in R with decreasing redshift.

(6) We find no correlation of Mg II doublet ratio with the equivalent
width ratios R and W(Fe II 2382)/W(Fe II 2600).  This clearly indicates
that the average column density of Fe II and column density ratios of
Fe II and Mg II are independent of Mg II column density.

(7) There is no strong tendency of Mg II equivalent width as well as Mg
II doublet ratio to show any change with redshift for systems with R
$>0.5$ . However there is a clear anti-correlation between Mg II
equivalent width and doublet ratio.

(8) For  systems with R$<0.5$, there is a clear tendency of equivalent
width and doublet ratio of the Mg II lines to increase with increasing
redshift. This indicates there are large number of low Mg II column
density clouds at higher redshift compared to lower redshift. The
doublet ratio  and equivalent widths of Mg II lines are not correlated
in these systems.

(9) Analysis of absorbers along the lines of sight searched for LLS and
Mg II absorption lines show no change in the number density of LLS
which do not show Mg II absorption. However the Mg II absorbers which
are nonLLS are seen only in the low redshifts.

(10) Nine individual metal line systems with $\tau_{\rm LLS}<3.0$ are
analysed with photoionization models to get an estimate of the
metallicity  and ionization parameters. Some of the absorbers in low
redshifts show metallicity almost that of sun. The required ionization
in most of the absorbers are less than 0.001.

The inferred ionization parameters for the high redshift absorbers are
high ( $\sim$0.01-0.001 at z$\simeq$1.5 (Srianand $\&$ Khare 1994);
$>3\times 10^{-3}$ at z$\sim$3.0 (Steidel 1990)). Thus the cloud will
have column density of Fe II compareble to that of Mg II only when it
is a damped Lyman alpha cloud.  Thus our results of more Mg II clouds
with R$<$0.5 at high z is consistent with high ionization
parameters at high redshift. The rough estimate of nonLLS Mg II systems
show at low redshifts 25$\%$ of the Mg II systems are not LLS (${\rm
dN\over dz} = 0.28$ compared to  0.96 for the whole sample).  This
together with the increase in the number of systems with R less than
0.5 with decreasing redshift suggest a decrease in the mean ionization
parameter. However there are LLS which do not show Mg II or Fe II
absorption even at low redshifts, with same number density as in  high
redshifts. Also available LLS data in low redshifts do not show any
change in $\tau_{\rm LLS}$ with redshift.

The number density per unit redshift of C IV systems decreases with
redshift for z$<2.0$, contrary to what is seen in very high redshifts
(Bahcall et al. 1993).  The values of ${\rm ({dN\over
dz})_{C\;IV}}\simeq0.87\pm0.43$ at average redshift 0.3 and
1.76$\pm$0.33 at an average redshift 1.50.  The ratio of number density
of C IV and Mg II  (defined as R1) at z = 1.5 is 1.6$\pm$0.57 and at z
= 0.3 the value of R1 is 1.16$\pm$0.73 (Bergeron et al.  1995). The two
values are consistent with one another within 1$\sigma$ uncertainty.
They found the systems with W(C IV)/W(Mg II) $<$ 1 contains 38$\%$ of
the absorbers at z = 0.53, whereas the fraction is only 17$\%$ at z =
1.7. As metagalactic radiation field  is lower at z =0.50 than at z =
1.7 by an order of magnitude, this suggests that if the gas is
photoionized by the background radiation, the mean gas density of the
high ionization systems falls with decreasing redshift. Their results
show even at low redshifts the high ionization absorbers are the
dominant over low ionization absorbers.

  Pettini et al.(1994) also noted that both distribution and typical
metallicity measured in the damped Lyman alpha clouds are more in line
with the values found in the halo. They also showed the observed
metallicities have a very large spread, and on a average the damped
Lyman alpha clouds at z$\simeq$2.0 have metallicity roughly 1/10th of
the solar value. Comparison of metallicity estimates obtained here with
the results of Pettini et al. (1994) suggests that on an average there
is an increase in the metallicity between z$\sim$2 to z$\sim$0.6.  Our
analysis of individual low optical depth LLS show that there are
atleast few systems which are enriched to solar metallicity at the look
back time corresponding to the formation epoch of sun in our Galaxy.
There are also few LLS which do not show Mg II or Fe II absorption
indicating the existance of low metallicity clouds also in the same
epoch.

Steidel $\&$ Dickinson (1994) showed based on the color distribution,
the Mg II line selected galaxies are undergoing constant star formation
in the whole redshift range studied (z$<$1.5). It is known that if the
star formation is continuous then one would expect that [O/Fe] would be
over solar, as well as [Mg/Fe] and [Si/Fe]. In addition [C/Fe] should
be close to solar where as [N/Fe] would be subsolar. If the star
formation in the absorbers are not continuous then the abundance
pattern will be different from those expected for continuous star
formation. Thus studying the abundance pattern in the redshift range
0.5 to 2.0, where most of the enrichment seems to have occured, will
give a clear picture of the chemical evolution of the galaxies.

Steidel $\&$ Dickinson (1994) also showed that the star formation rate
is not correlated with the properties of the halo. They concluded that
this result clearly rules out the possibility that the halo clouds are
originated from the disk of galaxies (fountain models), and the halo
gas may be due to satellite accreation. If the halo gas are due to
satellite accreation then the evolution of metallicity in the halo as
well as the disk will be very different with redshift. Studying the
metallicity distribution of damped Lyman alpha systems (which are
believed to be similar to galactic disks) and other metal line systems
( most probably originate in halos) will enable one to understand the
origin and evolution of halo gas.  It is also possible that the halo
clouds which are seen at z$\simeq$2.0 can undergo collision which
induces star formation and enrich the halo. The star formation also
helps in keeping the clouds in the halo for a long time. There are also
indications for the recent star formation in the halo of our galaxy.
Brown et al. (1989) have showed  few halo stars in our Galaxy have
flight time to reach its present z-position, if ejected from the disk,
much more than their evolutionary age, implying that they must have
formed in the halo. Future detail studies with extended UV data in
the low redshift range can discriminate between various alternatives.

\eject

\eject

\centerline{\bf Figure captions}
\vskip0.1in
\noindent {\bf Figure 1: } Relation between the N(Mg II) and N(Fe II),
obtained based on single cloud curve of growth, required to produce
different values of R for different velocity dispersion parameters (30,
40, 50, 60, 70 kms$^{-1}$).

\vskip0.1in
\noindent {\bf Figure 2: } Results of photoionzation models
for different values of ionization parameter and metallicity ( Z =
0.001, 0.01, 0.1, 0.5 Z$_\odot$).

\vskip0.1in
\noindent {\bf Figure 3: } Plot showing R as a function of redshift.
The filled squares represent the upper limits.

\vskip0.1in
\noindent {\bf Figure 4: } Histogram of R in two different redshift
bins.
\eject
\begin{table}
\centerline{\bf Table 1}
\centerline{\bf Results of photoionization models}
\begin{center}
\begin{tabular}{ccccc}
\tableline
\tableline
\multicolumn {2}{c}{}&\multicolumn {2}{c}{Minumum log N(H I) for}&
\multicolumn {1}{c}{}\\
\multicolumn {1}{c}{log $\Gamma$}&
\multicolumn {1}{c}{Z/Z$_\odot$}&
\multicolumn {1}{c}{N(Fe II)$>10^{13}$ }&
\multicolumn {1}{c}{N(Fe II)$\ge$N(Mg II)}&
\multicolumn {1}{c}{log N(Mg I)}\\
\tableline
-2.0 &0.01&19.00&21.00&12.56\\
     &0.10&18.40&20.60&13.40\\
     &0.50&18.00&20.40&13.87\\
-2.5 &0.01&19.00&20.40&12.33\\
     &0.10&18.20&20.40&13.32\\
     &0.50&17.60&20.00&13.72\\
-3.0 &0.01&19.00&20.00&12.04\\
     &0.10&18.00&19.80&12.93\\
     &0.50&17.20&19.60&13.42\\
-3.5 &0.01&19.20&19.60&11.70\\
     &0.10&17.60&19.40&12.58\\
     &0.50&16.80&19.00&13.00\\
-4.0 &0.01&19.20&19.00&11.14\\
     &0.10&17.60&19.00&12.13\\
     &0.50&16.80&18.20&12.34\\
\tableline
\end{tabular}
\end{center}
\end{table}
\vfill\eject
\newpage

\begin{table}
\centerline{\bf Table 2}
\centerline{\bf Results of Maximum likelihood analysis}
\begin{center}
\begin{tabular}{cccccc}
\tableline
\tableline
\multicolumn{1}{c}{${\rm {W_r}^{min}}$}&\multicolumn{1}{c}{Number}&
\multicolumn{1}{c}{$<{\rm z}>$}&\multicolumn{1}{c}{N(z)}  &
\multicolumn{1}{c}{$\gamma\pm{\rm d}\gamma$}&\multicolumn{1}{c}{${\rm
W_*\pm dW_*}$}\\
\tableline
\multicolumn{6}{c}{Steidel$\&$Sargent 92}\\
0.30&111&1.12&0.97&0.78$\pm$0.42&0.57$\pm$0.06\\
0.60&67&1.17&0.52&1.02$\pm$0.53&0.66$\pm$0.08\\
1.00&36&1.31&0.27&2.24$\pm$0.76&0.66$\pm$0.11\\
\tableline
\multicolumn{6}{c}{for R$<$0.5 systems}\\
0.30&43&1.29&0.38&2.27$\pm$0.71&0.74$\pm$0.11\\
0.60&39&1.41&0.31&3.09$\pm$0.78&0.69$\pm$0.11\\
1.00&22&1.49&0.17&4.24$\pm$1.13&0.68$\pm$0.15\\
\tableline
\multicolumn{6}{c}{for Fe II lines}\\
0.30&32&1.18&0.28&1.14$\pm$0.77&0.44$\pm$0.07\\
0.60&20&1.24&0.15&1.60$\pm$0.99&0.36$\pm$0.08\\
1.00&9 &1.09&0.06&0.43$\pm$1.38&0.21$\pm$0.06\\
\tableline
\end{tabular}
\end{center}
\end{table}

\begin{table}
\centerline{\bf Table 3}
\centerline{\bf Results of Spearman correlation test}
\begin{center}
\begin{tabular}{ccccc}
\tableline
\tableline
{Test variables}&{$\rho$}&$\rho/\sigma$&significance\\
\tableline
W(Fe II),z&0.11&0.76&0.450\\
${\rm W(Fe II 2382)\over W(Fe II 2600)}$,z&0.07&0.48&0.625\\
R,z&-0.27&-1.91&0.055\\
R,dr(Mg II)&-0.09&-0.69&0.515\\
${\rm W(Fe II 2382)\over W(Fe II 2600)}$,dr(Mg II)&0.15&1.01&0.286\\
\tableline
&\multicolumn {2}{c}{when R$<$0.5}& \\
W(Mg II),z&0.37&2.72&0.005\\
dr(Mg II),z&0.25&1.83&0.067\\
dr(Mg II),W(Mg II)&-1.63&-1.19&0.239\\
\tableline
&\multicolumn {2}{c}{when R$>$0.5}& \\
W(Mg II),z&0.09&0.45&0.657\\
dr(Mg II),z&-0.26&-1.32&0.189\\
dr(Mg II),z&-0.51&-2.58&0.007\\
\tableline
\end{tabular}
\end{center}
\end{table}

\end{document}